\documentstyle[aps,epsf]{revtex}
\begin{document}
\begin{title}
{Two-Proton Correlations Relative To The Reaction Plane}
\end{title}

\author{Sergei Y. Panitkin$^{1}$ for the E877
Collaboration$^{2}$} 

\address{$^{1}$Physics Department, Kent State University, Kent, OH 44242-0001\\
  $^{2}$BNL - GSI - INEL - McGill Univ. - Univ. of Pittsburgh - SUNY
  Stony Brook - Univ. of Sao Paulo - Wayne State Univ.} 
\maketitle
\section{INTRODUCTION}
Studies of the collective flow of hadrons in all of its forms -
directed, elliptic, radial is an important direction in the
understanding of the physics of heavy-ion
collisions~\cite{review}. The asymmetries observed in
azimuthal distributions in momentum space 
imply underlying asymmetries in configuration space. But the
experimental information about spatial properties of the flowing
nuclear systems is at best sparse 
and is mostly obtained with pion measurements. The E877 collaboration
reported~\cite{pipi_qm95} that 
the parameters of the pion source created in Au+Au collisions at 10.8
AGeV/c appear to be different for particles emitted at different angles
relative to the reaction plane. 
The information about space-time properties of the proton effective
source involved in directed flow is totally absent. 
Since nucleons are the main carriers of the directed flow in nucleus
nucleus collisions at the AGS energies, it is interesting to check
whether the parameters 
of the nucleon source, probed with the help of two-proton correlations,
exhibit any dependencies or asymmetries related to the reaction plane
orientation. 
In this paper we present the preliminary results of the first study of
the proton correlation function's dependence on the orientation of the
reaction plane.
\section{EXPERIMENTAL SETUP}
The measurements are made by the E877 collaboration at the BNL AGS
during the 1994 run. The details of the E877 experimental setup can be
found elsewhere~\cite{Barrette94_1,Barrette97_1,Barrette97_2}.
Here we will only briefly describe its features relevant to the
present analysis.  A set of two high granularity sampling 
calorimeters positioned around the target area provide an event
by event measurement of the transverse energy ($E_{T}$) production in
the interval of pseudo rapidity -2.0$<\eta<$4.7 .
A high-resolution forward magnetic spectrometer  has been used for
detection and identification of charged particles.
Simultaneous measurements of the particle rigidity and time of flight in
the spectrometer provide particle identification up to the momentum of
the beam. The momentum resolution, found
from Monte Carlo studies, is better than 3\% over the entire
momentum range. The momentum resolution is in good agreement with the
momentum dependence of the measured width of the charged particle
mass peaks. Corresponding relative invariant momentum resolution is
estimated to be of order of 7 MeV/c for Q$_{inv} \le 40$MeV/c. In the 
same interval of Q$_{inv}$ the average rapidity of the protons is about
2.6 and the average transverse momentum about 0.3 GeV/c. Note that beam
rapidity is about 3.1.
\section{METHODS}
\subsection{Reaction Plane Determination}
\indent The reaction plane is a fundamental plane of symmetry of the
reaction. Knowledge of it is clearly important for the two-particle 
correlation 
analysis which attempts to extract information about the spatial-temporal
structure of the system created in the collision.
Broad pseudo rapidity coverage and high segmentation of the E877 calorimeters 
provide the experiment with the information about the reaction
plane. The method of an event by event reaction plane   
determination without particle identification is based on the idea of 
performing a Fourier analysis on the transverse energy deposited in a
fixed pseudo-rapidity window~\cite{Barrette94_1,YZhang96}.
The Fourier coefficients of the expansion are given by:
\begin{equation}
a_{n} = \frac{\sum_{i} \epsilon_{i} cos(n \phi_{i})}{\sum_{i} \epsilon_{i}},
\hspace{0.2in}
b_{n} = \frac{\sum_{i} \epsilon_{i} sin(n \phi_{i})}{\sum_{i} \epsilon_{i}}
\end{equation}
\noindent
where $\phi_{i}$ is the azimuthal angle of the $i$-th detector cell
and $\epsilon_{i}$ is the energy deposited in that cell.  The
n=1 Fourier coefficients hold information regarding the azimuthal
orientation of the energy flow in the collision.
The vector $\overrightarrow{flow}=(a_{1}, b_1)$ points in the
direction where most of the transverse energy is carried by the
particles emitted from the collision zone.  
To put it another way, this vector is related to the azimuthal
orientation of the impact parameter. Therefore, the azimuthal angle of
the reaction plane may be determined as:   
\begin{equation}
\psi_{reac} = tan^{-1}\left( \frac{b_{1}}{a_{1}}\right) 
\end{equation}
From the naive geometrical considerations one would expect that the
proton source will be at its 
most symmetric configuration at impact parameter zero and maximum
asymmetry 
will be at maximal impact parameter. So in order to study the reaction
plane dependence one would need to use the most peripheral event
sample. The interval of centralities chosen for our analysis
represents a compromise between the necessity of a high
statistics sample and the desire to utilize the least central
collisions with the best achievable reaction plane resolution as well
as a need for a fairly narrow interval of impact parameter. 
Events with centrality within the interval from $9\%$ to 
$6\%$ of total geometrical cross section are selected for this
analysis. The reaction plane resolution at this centrality interval is
about 40 degrees and is close to best achievable by the E877 setup. 
\subsection{Proton Correlations}
Two-proton correlations are due to the
attractive strong and 
repulsive Coulomb final state interactions and are also influenced by
the effects of quantum statistics which requires an antisymmetrization
of the two-proton wave function.
Coulomb repulsion, together with antisymmetrization, decreases the
probability of detection of pairs with relative momentum close to zero,
while the strong interaction increases this probability.
The interplay of these effects lead to a characteristic ``dip+bump''
shape of the correlation function. 
The height of the peak of the correlation function can be related to the
space-time parameters of the emitting source. 
It has been shown~\cite{Koonin77,Lednicky82} that, for 
simple static sources, the height, by which the peak deviates from
unity,  scales approximately inversely proportional to the source
volume.\\ 
\indent The experimental correlation function C$_2$ is defined as:
\begin{eqnarray} C_2(P_{cm}) =
\frac{N_{tr}(P_{cm})}{N_{bk}(P_{cm})} \end{eqnarray}
                                                                        \noindent{where $P_{cm}$ is a momentum difference in the pair's
rest frame. For pairs of the same mass $P_{cm}$ is equal to the
four-momentum difference:  

\begin{eqnarray}                                                     
P_{cm} = Q_{inv} =  \sqrt{-(p_1^{\mu} - p_2^{\mu})^2}              
\end{eqnarray}
The conventional variable for two-proton
correlation studies is $q_{inv}=\frac{1}{2} Q_{inv}$.} 
N$_{tr}$ and N$_{bk}$ in (1) are the ``true'' and ``background''
two-particle 
distributions obtained by taking particles from the same and different
events, respectively.
A condition that the two particles do not share the same slat of the
TOF hodoscope has been imposed on pairs from both distributions.
In order to account for the distortions of the correlation function
introduced by the two-track reconstruction inefficiency in high
multiplicity events, cuts on
separation of two tracks in the drift chambers are introduced.
Monte Carlo studies show that the cuts effectively suppressed these
distortions.In addition to standard cuts on pairs, more selection
criteria are introduced. Proton pairs are 
subdivided into several groups depending on the azimuthal 
angle of emission relative to the reaction plane angle. For true
pairs, both protons in a pair were required to have an angle of
emission relative to the reaction plane to be within a
certain interval. 

\begin{figure}
\begin{center}
{\epsffile{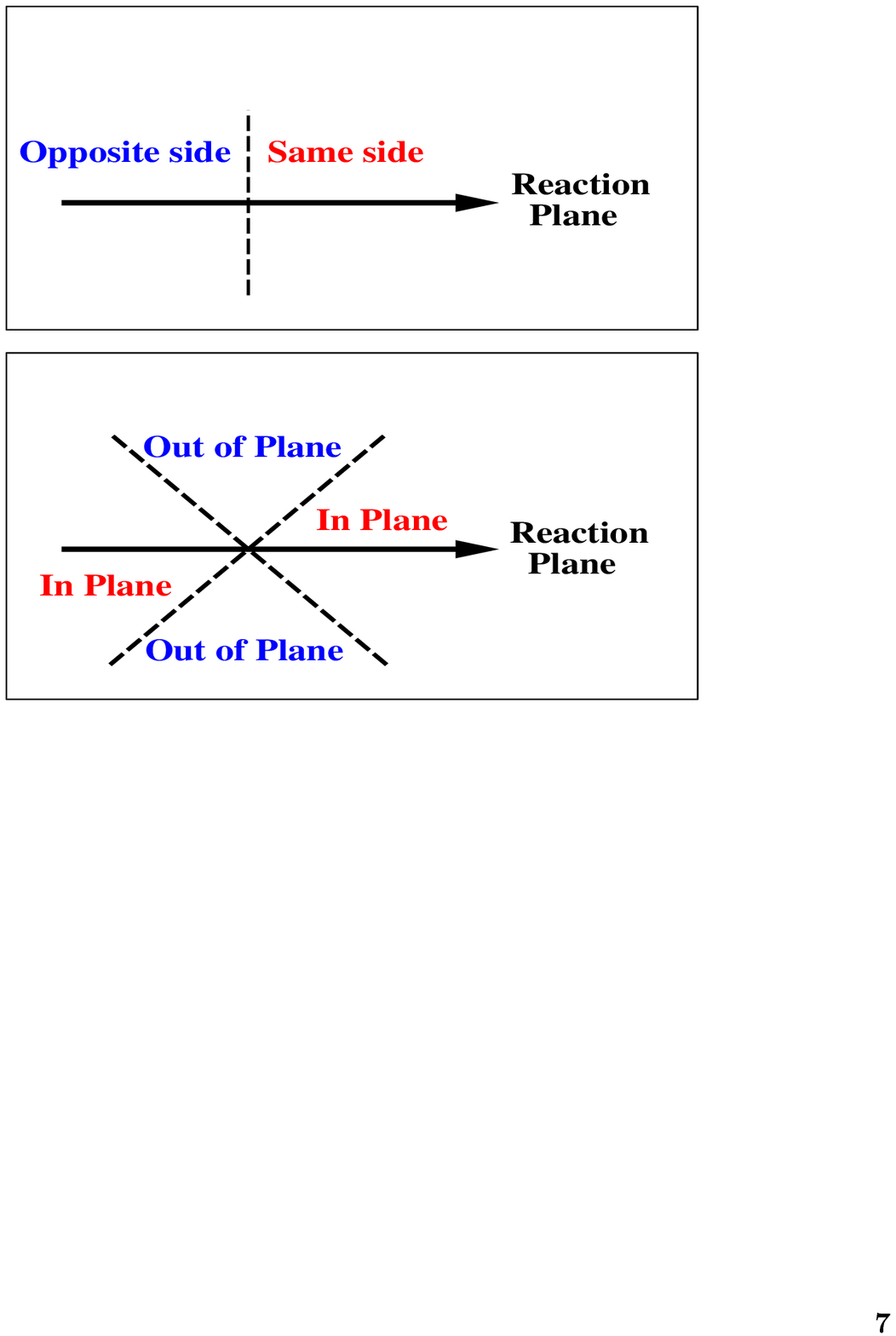}}
\end{center}
\caption{\footnotesize Schematic diagrams of the applied cuts on the
orientation 
relative to the reaction plane. Upper panel: the ``same'' and
``opposite'' orientations relative to the reaction plane; Lower panel:
``in-plane'' and ``out-of-plane'' cuts.} 
\label{Fig:cartoon}
\end{figure}

Figure~\ref{Fig:cartoon} presents a visualization of
the cuts on the angles relative to the reaction plane. In the mixed
events pool, every particle has its angle 
relative to the reaction plane stored as a part of its data
structure. Subsequently, during the event 
mixing procedure, particles that make pairs are required to
satisfy the same cuts on the angle relative to the reaction plane as
were imposed on the true pairs. 
\section{RESULTS AND DISCUSSION}
\subsubsection{Opposite Side vs Same Side} 
This is the simplest possible cut. Pairs are divided into two subsets
depending on the direction of the particle's emission angle relative
to the reaction plane. The ``same side'' subset is defined in the
following way: each 
proton of a pair is required to have a positive cosine of the angle
relative to the direction of the reaction plane. The ``same side''
subset yields 753k proton pairs. The ``opposite side''
subset is comprised of pairs with a negative cosine of the angle
relative to the reaction plane. It has 568k 
pairs.

\begin{figure}
\begin{center}

\epsfxsize =0pt
\epsffile[0 200 300 700]{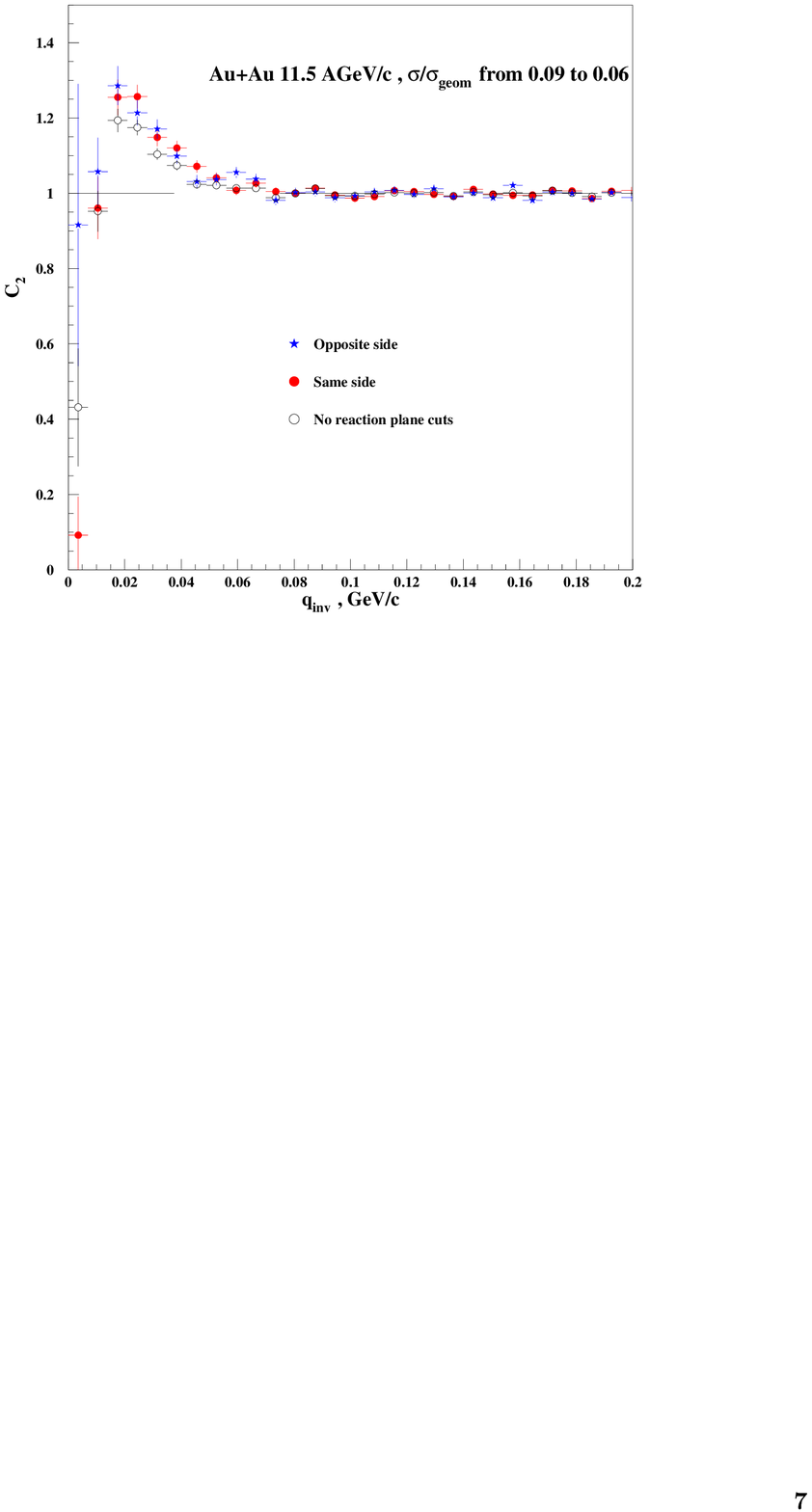}
\end{center}
\caption{ Two-proton correlation functions for the ``same'' and
``opposite'' orientations relative to the reaction plane.} 
\label{Fig:same_opposite}
\end{figure}

Figure~\ref{Fig:same_opposite} shows the
two-proton correlation functions for these two subsets.
One can conclude from the figure that there is no significant
differences between the ``same side'' and ``opposite side''
correlation functions.
\subsubsection{In-Plane vs Out-of-Plane}
Another set of cuts is intended to
probe the differences between particles emitted close to the reaction
plane and particles emitted out of the reaction plane. The
``in-plane'' subset 
is defined as having the absolute value of the cosine of an angle between
the particle and the reaction plane direction smaller than
cos($\pi$/4). Both particles of a pair are required to be in the same
quadrant. In the case when the cosine is greater than cos($\pi$/4)
pairs are labeled as emitted ``out of plane''. The ``in-plane''
sample yields 596k events, the ``out-of-plane'', 582k
pairs. 

\begin{figure}
\begin{center}
\epsfxsize = 0 pt
\epsffile[0 200 300 700]{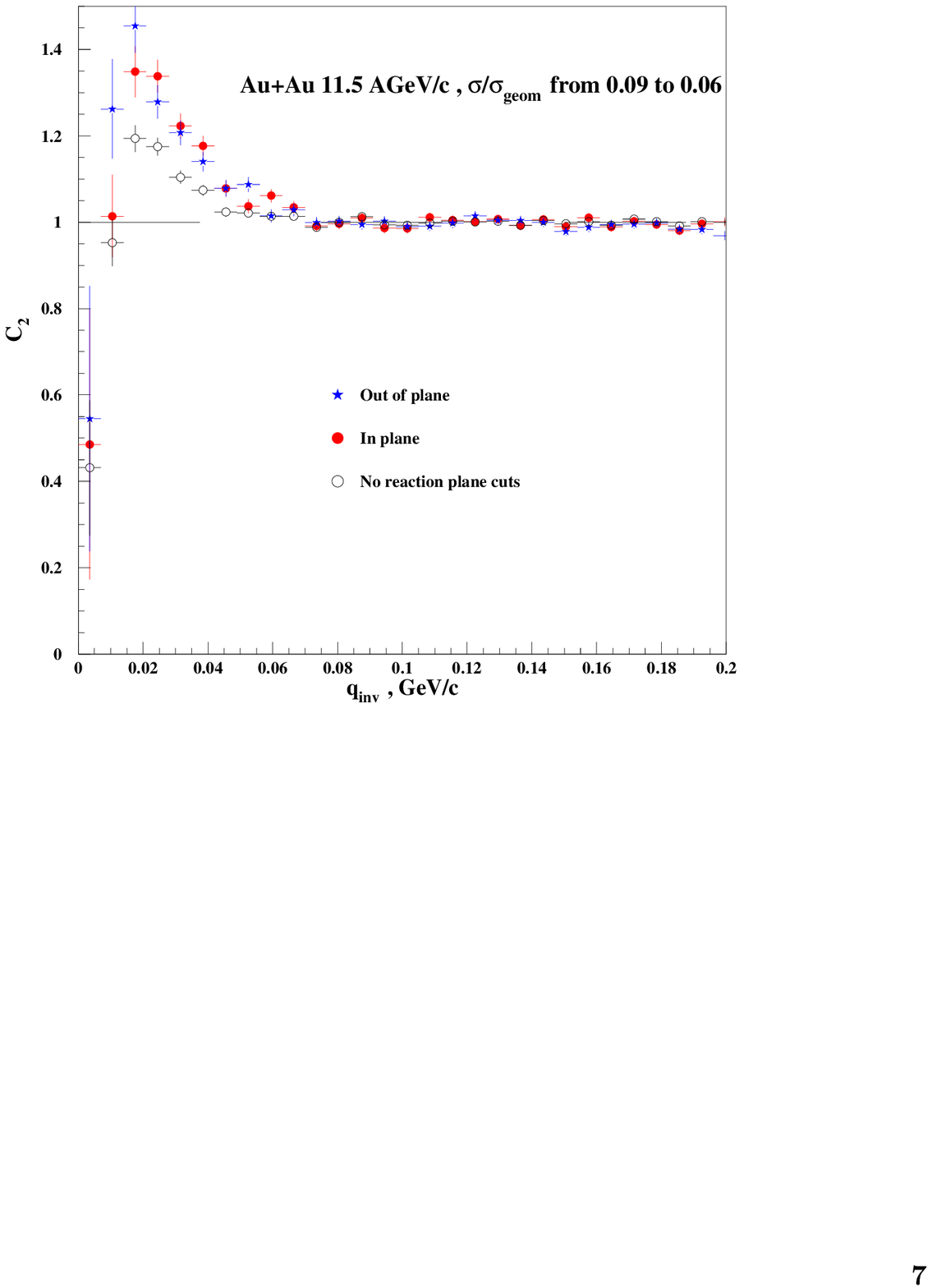}
\end{center}

\caption{Two-proton correlation functions for the ``in-plane'' and
``out-of-plane'' orientations of proton pairs relative to the reaction
plane. The correlation function obtained without reaction plane cuts is
also shown with open circles.}  
\label{Fig:inplane_out_of_plane}
\end{figure}

Figure~\ref{Fig:inplane_out_of_plane} shows the two-proton 
correlation functions for these ``in-plane'' and ``out-of-plane''
subsamples. It can be seen from the figure that 
there are no statistically significant differences between
``in-plane'' and ``out-of-plane'' correlation functions. However, both
subsamples yielded 
correlation functions with peaks which are significantly higher than the
correlation function for the case when no cut on the angle relative to
reaction plane is imposed. Similar trend can be also observed on
Figure~\ref{Fig:same_opposite} though it is not as pronounced.
This behavior may be explained within the picture suggested by
Voloshin~\cite{Voloshin97_1}- a transverse radial expansion of
the proton source which moves sideward in reaction plane. This model
seems to be able to explain the behavior of some rather detailed
features of the proton and pion directed
flow~\cite{Barrette97_2,Voloshin98}. 
In this framework the observed behavior of the proton correlation
functions may be explained by a correlation between position and
direction of emission induced by transverse expansion of the flowing
system. Angular cuts select particles emitted in the 
same direction in the azimuthal plane (lower relative momentum) which
are likely to be on the same ``side'' of the expanding system (smaller
spatial separation). Both effects ``amplify'' the peak of the
correlation function\footnote{For details of model analysis of the
influence of radial transverse expansion on two-proton correlation
function see~\cite{Alladin}.}  
\section{Model Calculations}
\begin{figure}
\begin{center}
\vspace{12cm}
\epsfxsize = 0 in 
\epsffile[0 200 300 700]{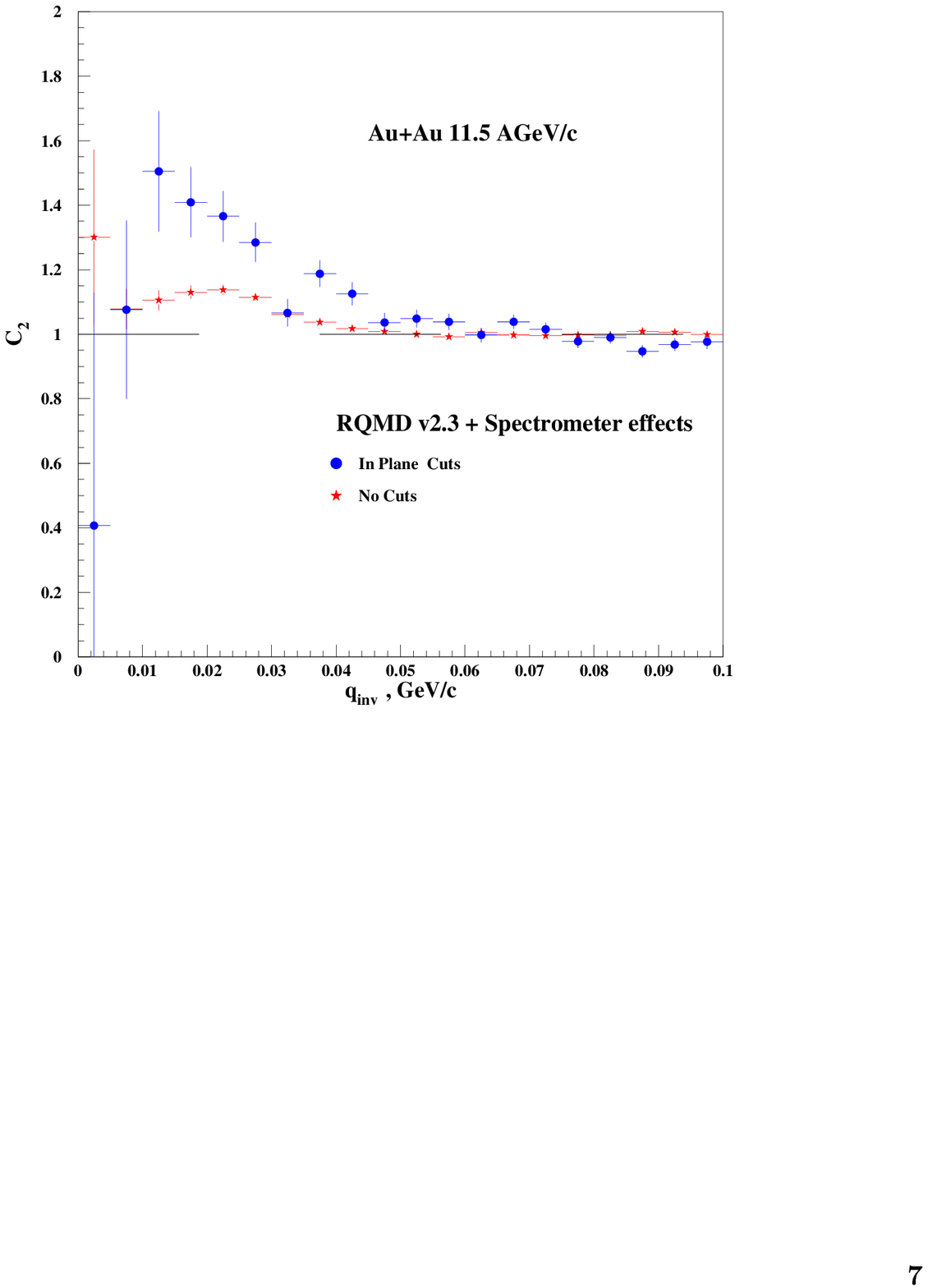}
\end{center}
\caption{Model two-proton correlation functions for the ``in-plane''
(solid circles) orientation of proton pairs relative to the reaction 
plane. The correlation function obtained without reaction plane cuts
is shown with stars.}  
\label{Fig:rqmd_inplane_all}
\end{figure}

\begin{figure}
\begin{center}
\epsfxsize = 0  in 
\epsffile[0 200 300 700]{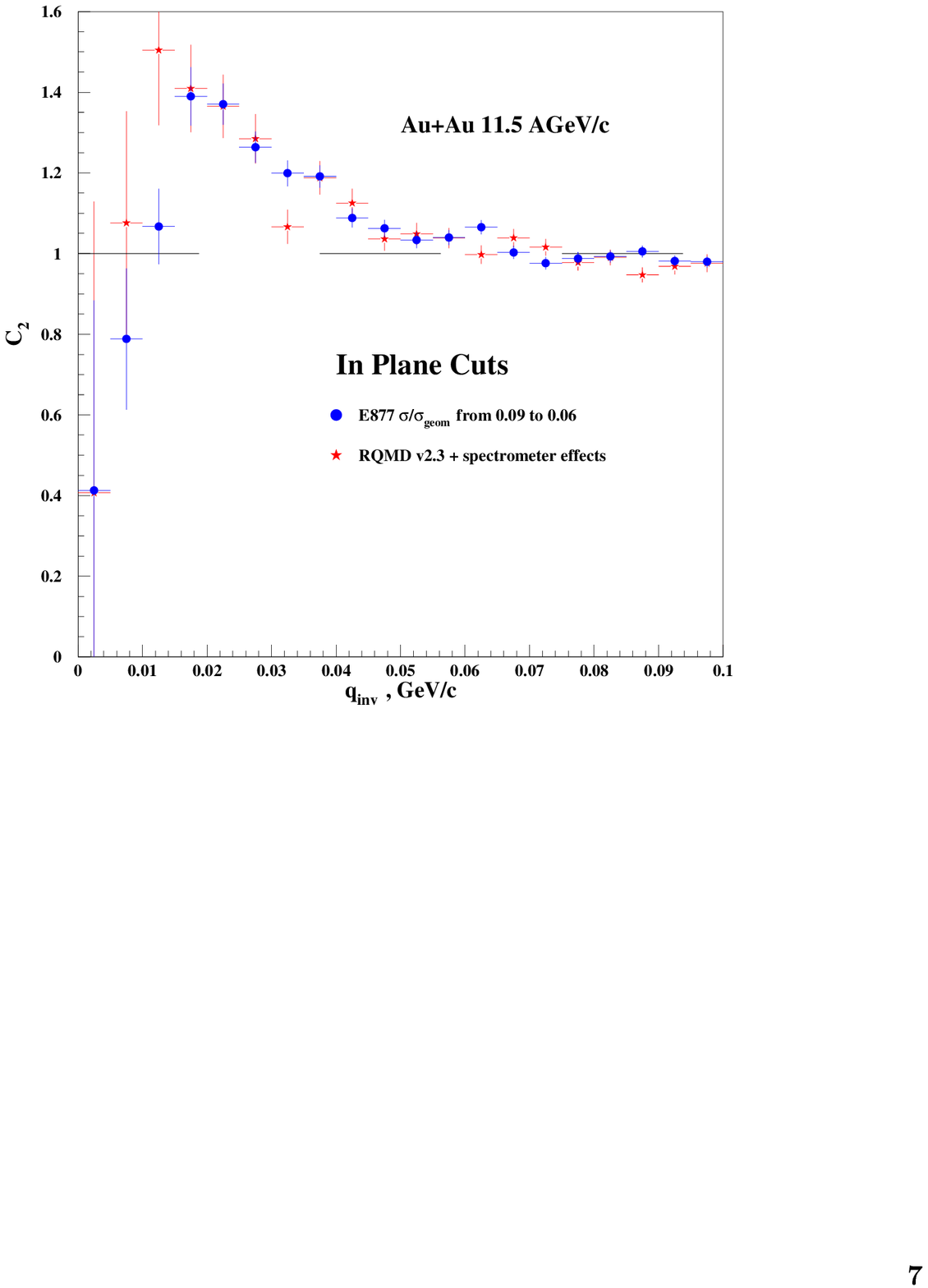}
\end{center}
\caption{Model correlation function for the ``in-plane'' cut compared
to with the measured one.}   
\label{Fig:rqmd_data_inplane}
\end{figure}

In order to extract more physical information from the measured
correlation function, we carried out a 
study using the phase space produced by the
event generator RQMD(v2.3)~\cite{rqmd2,rqmd3}. This model
describes 
classical propagation of the particles, together with quantum effects of
stochastic scattering and Pauli blocking. It includes color strings
and ropes, baryon and meson resonances, as well as finite
formation time for created particles. It has been successful in the
description of many 
features of relativistic heavy-ion collisions. In this model
a particle's freeze-out position is defined as a point of the last
strong interaction. The structure of our approach is as follows: 
by taking the freeze-out phase-space distribution generated by RQMD
and propagating the particles through the 
experimental acceptance, accounting for the resolution of the
detectors, a subset of the phase-space points is
obtained. Then the Koonin-Pratt
method\cite{Pratt90,Pratt94} 
is used to construct the proton-proton correlation function. This
method provides a description of the final state interactions between
two protons and antisymmetrization of their relative wave
function.\\
\indent Simulated pairs are subjected to the same cuts as real
data. Angular cuts are performed relative to the impact parameter
direction of the RQMD events. 
As an example\footnote{For brevity sake we will limit our discussion
to the ``in-plane'' cut. The ``out of plane'' one shows similar
behaviour.}
Figure~\ref{Fig:rqmd_inplane_all} shows the results of 
the calculations with ``in-plane'' cuts together with the calculated
correlation 
function without the reaction plane cuts. It is clear that the model
correlation functions exhibit behavior qualitatively similar to the
one observed in data. 
Note that correlation functions on
Figures~\ref{Fig:rqmd_inplane_all} and ~\ref{Fig:inplane_out_of_plane}
have different binning. Figure~\ref{Fig:rqmd_data_inplane} shows
calculated and measured correlation functions for the proton pairs
emitted ``in-plane''. One can see that the agreement between data and
model calculations is rather good. Assuming that RQMD simulates the
freeze-out phase-space and configuration space distributions
correctly, one can try to gain  some insight from studying the
parameters of the proton source in RQMD. Statistical parameters of the
distributions in relative coordinates and emission times are listed in
the Table~\ref{Tab:rqmd}. An interesting feature that can be seen from
the Table~\ref{Tab:rqmd} is that the difference
between parameters of the two subsamples is less then half
fermi. A case of a chaotic static source can not account for
an observed difference in peak height of the correlation functions
with and without angular cut. The difference is due to a
momentum-position correlation of the model source.
 
\begin{table}
\begin{center}
\begin{tabular}{lrlrlrlrl}\hline
 RMS      &  $\Delta X$ & &  $\Delta Y$ & &  $\Delta
Z$ & &  $\Delta c\tau$ & \\\hline
 In-Plane Cut&  5.44     &  &   5.36    & &    16.9
 & &    18.36     &   \\ 
 All pairs     & 5.80      & &  5.83     & &    16.4
& &    18.66  & \\\hline 
 \end{tabular}
\end{center}
\caption{Width of the distribution of relative space and time
coordinates for proton pair with and without in-plane cuts. All source
parameters are in fermi. See description in the text.}
\label{Tab:rqmd}
\end{table}
 
\section{SUMMARY}
In summary, we have investigated dependence of the two-proton
correlation function on the orientation relative to the reaction
plane in Au+Au collisions at 11.5 AGeV/c.
We conclude that (at currently achieved sensitivity) the two-proton
correlation function does not exhibit significant differences when a
proton source is probed from different sides of the reaction plane,
which suggest an azimuthal symmetry of the source. We also conclude
that the observed difference in the correlation functions with and
without cuts on the orientation of pairs relative to the reaction 
plane is likely to be linked to the momentum-space correlations
induced during expansion of the proton source. 
Model calculations based on RQMD and Koonin-Pratt formalism agree with
data fairly well.
\section{Acknowledgments}
Experiment E877 is supported in part by the US DoE, the NSF, the
Canadian NSERC and CNPq Brasil.

\pagebreak
\pagebreak
\vspace{5cm}

\end{document}